\documentclass[prb,reprint,nofootinbib]{revtex4-1}
\bibpunct{[}{]}{;}{n}{}{}

\pdfoutput=1
\usepackage[caption=false]{subfig}

\usepackage{graphicx}
\usepackage{comment}
\usepackage{amsmath}
\usepackage{amssymb}
\usepackage{amsfonts}
\usepackage{bbm}
\usepackage{braket}
\usepackage{verbatim}
\usepackage[hidelinks]{hyperref}
\usepackage{ulem}
\usepackage{color, colortbl}
\usepackage{multirow}
\definecolor{LightCyan}{rgb}{1,0.5,0.5}
\hypersetup{
  colorlinks   = true, 
  urlcolor     = blue, 
  linkcolor    = blue, 
  citecolor   = blue 
}

\usepackage{color}
\usepackage{tikz}

\newcommand{\be}{\begin{equation}}
\newcommand{\ee}{\end{equation}}

\newcommand{\gammabeta}{\{\gamma, \beta \}_\text{tree}}

\newcommand{\gb}[0]{{\gamma, \beta}}
\newcommand{\gbcostof}[1]{F_p^{#1}(\gb)}

\begin{document}

\title{The fixed angle conjecture for QAOA on regular MaxCut graphs}
\author{Jonathan Wurtz}
\affiliation{Department of Physics and Astronomy, Tufts University, Medford, Massachusetts 02155, USA}
\email[Corresponding author: ] {jonathan.wurtz@tufts.edu}
\author{Danylo Lykov}
\affiliation{Argonne National Laboratory, Lemont, Illinois 60439, USA}

\date{\today}

\begin{abstract}
The quantum approximate optimization algorithm (QAOA) is a near-term combinatorial optimization algorithm suitable for noisy quantum devices. However, little is known about performance guarantees for $p>2$. A recent work \cite{Wurtz_guarantee} computing MaxCut performance guarantees for 3-regular graphs conjectures that any $d$-regular graph evaluated at particular fixed angles has an approximation ratio greater than some worst-case guarantee. In this work, we provide numerical evidence for this fixed angle conjecture for $p<12$. We compute and provide these angles via numerical optimization and tensor networks. These fixed angles serve for an optimization-free version of QAOA, and have universally good performance on any 3 regular graph. Heuristic evidence is presented for the fixed angle conjecture on graph ensembles, which suggests that these fixed angles are ``close" to global optimum. Under the fixed angle conjecture, QAOA has a larger performance guarantee than the Goemans Williamson algorithm on 3-regular graphs for $p\geq 11$.
\end{abstract}

\maketitle

\section{Introduction}

Near-term quantum computers have the potential for advantage in the field of combinatorial optimization. Potentially, an algorithm running on a small noisy quantum device may soon exhibit quantum advantage by providing better approximate to combinatorial problems than the best classical approximate solver \cite{Bravyi2018advantage}. One particular class of potential algorithms are variational quantum algorithms (VQA) \cite{VQA_overview}, which use a hybrid quantum-classical loop to optimize ansatz wavefunctions that encode solutions to combinatorial problems.

One particular ansatz choice for VQA is the quantum approximate optimization algorithm (QAOA) \cite{farhi2014quantum}, which is generated using $p$ rounds of unitaries alternating between some ``mixing" unitary and objective function unitary. The ansatz is a function of $2p$ parameters $\{\gamma,\beta\}$, which are optimized through repeated query of a classical optimizer to a digital quantum device.

While it is known that the QAOA converges to the exact result for $p\to\infty$ consistent with the adiabatic theorem \cite{farhi2014quantum,wurtz2021counterdiabaticity}, less is known about the guaranteed performance of the algorithm at low depth. In the original introduction of Farhi et.~al \cite{farhi2014quantum}, the performance was guaranteed to be at least $0.693$ for $p=1$ and 3 regular graphs, and in a later work \cite{Wurtz_guarantee}, the performance was guaranteed to be at least $0.7559$ for $p=2$ and 3 regular graphs. The same work observed that worst-case graphs have no small cycles, and conjectured that the same holds for larger $p$.

In this work, we provide performance guarantees under this conjecture for regular graphs for $p\leq 11$, and show that for $p\geq 11$ the QAOA has a larger performance guarantee on 3 regular graphs than the best general purpose classical solver.

Additionally, it has been observed that optimal QAOA angles concentrate around typical values \cite{brandao2018fixed,galda2021transferability}, suggesting that QAOA may bypass the quantum-classical variational optimization step, or pre-compute the angles using a tensor network quantum circuit simulator \cite{streif2019training}. In this work, we make this observation even more simple by demonstrating that there exist a set of fixed angles that are ``universally good". These fixed angles, when evaluated on any regular graph of fixed edge weights, will return approximation ratios larger than some guarantee and very close to the exact maximal approximation ratio. These fixed angles allow QAOA to completely bypass any variational optimization step, potentially increasing execution times by a factor of 100 - 1000$\times$.

\begin{figure*}
\centering
\includegraphics{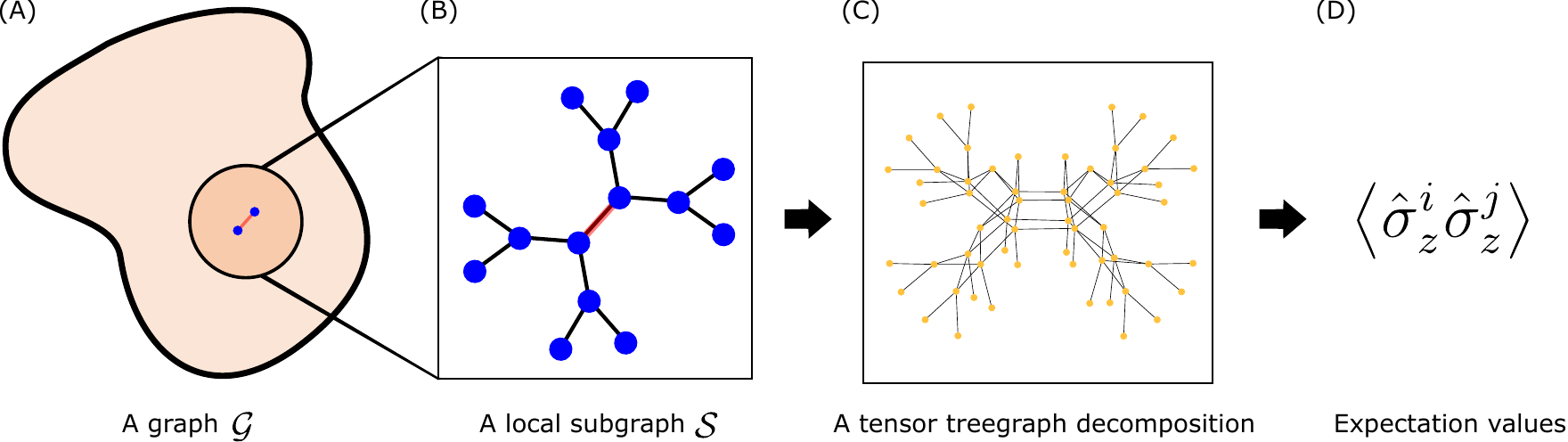}
\caption{A sketch of how to compute QAOA expectation values. Given some graph $\mathcal G$ of edges and vertices \textbf{(A)}, QAOA will only ``see" local structure within $p$ steps of an edge. In this way, a graph can be decomposed into $|E|$ subgraphs $\mathcal S$ \textbf{(B)}, one for each edge, and expectation values be computed independently \textbf{(D)}. One method of computing expectation values used in this work are tensor networks \textbf{(C)}, which efficiently decompose the local QAOA state into a tensor network for efficient computation.}
\label{fig:tree_subgraph}
\end{figure*}

\quad

In this work, we focus on the \texttt{NP-complete} combinatorial optimization problem MaxCut \cite{garey1976}. Given a graph $\mathcal G$ of edges $E$ and vertices $V$, a MaxCut algorithm strives to find a bipartition of vertices $\{A,\,V\backslash A\}$ such that a maximal number of edges $E$ have one vertex in each bipartition. In other words, a MaxCut strives to ``cut" the maximal number of edges via a bipartition.

A graph of $N=|V|$ vertices and $M=|E|$ edges can be encoded into an objective function $C$ over $N$ qubits and $M$ clauses, as follows

\begin{equation}
\hat C = \frac{1}{2}\sum_{\langle ij\rangle \in E}(1 - \hat \sigma_z^i\hat \sigma_z^j).
\end{equation}

The eigenstates of $\hat C$ are bipartitions, with eigenvalues counting the number of cut edges. The partition is assigned by measuring the ansatz state in the $Z$ basis.

The QAOA strives to find optimal parameters which maximize the expectation value of the objective function with respect to some ansatz state. The QAOA ansatz is defined as a function of $2p$ variables $\{\gamma,\beta\}$ over $p$ rounds of repeated unitaries

\begin{equation}\label{eq:QAOA_ansatz}
|\gamma,\beta\rangle = e^{-i\beta_p \hat B}e^{-i\gamma_p \hat C}(\cdots)e^{-i\beta_1 \hat B}e^{-i\gamma_1 \hat C}|+\rangle
\end{equation}
where ellipsis represent the $p$ actions of objective function $\hat C$ and mixing term $\hat B=\sum_i\hat \sigma_x^i$. The initial state $|+\rangle$ is the equally weighted superposition state and maximal eigenstate of $\hat B$.

The QAOA is a hybrid algorithm, in the sense that it includes both a quantum and a classical device. Through repeated query to a small quantum device to evaluate expectation values $F_p^{\mathcal G}(\gamma,\beta) = \langle \gamma,\beta|\hat C|\gamma,\beta\rangle$, a classical device optimizes the $2p$ angles $\{\gamma,\beta\}$ to provide optimal MaxCut solutions to a graph.

The performance of the algorithm is characterized by the approximation ratio, which is the ratio between the optimized value $F_p^{\mathcal G}(\gamma,\beta)$ and the MaxCut value

\begin{equation}
C_p^{\mathcal G} = \;\underset{\gamma,\beta\;}{\texttt{MAX:}}\; \frac{F_p^{\mathcal G}(\gamma,\beta)}{C_\text{max}^{\mathcal G}}.
\end{equation}

The approximation ratio ranges between $0$ and $1$. A larger indicates better performance, and a value of 1 indicates the exact solution. Given a class of graphs and fixed value $p$, this approximation ratio may be guaranteed to be above some performance guarantee. For 3 regular graphs, $C_1\geq 0.692$ \cite{farhi2014quantum}, and $C_2\geq 0.7559$ \cite{Wurtz_guarantee}. For $p\to\infty$ the approximation ratio converges to 1 in accordance with the adiabatic theorem \cite{farhi2014quantum,wurtz2021counterdiabaticity}.

Critically important to the evaluation of the performance is the fact that QAOA is a local algorithm \cite{gamarnik2013limits,farhi2020quantum,farhi2020quantum2,barak2021classical}. Given $p$ steps, a vertex can only be correlated with other vertices within a distance $\leq p$. This is because the QAOA has a ``lightcone" of interaction, which is clear to see in the Heisenberg picture (See for example, \cite{Wurtz_guarantee} Eq. 7). The expectation value of the cost function is a sum of clauses
\begin{align}
\gbcostof{\mathcal G} &= \braket{\gb| \hat C | \gb}\nonumber
\\
&= \bra{\gb} \sum_{\langle ij\rangle\in E}\frac{1}{2}(1-\hat \sigma_z^i\hat \sigma_z^j)\ket{\gb}\nonumber\\
&\equiv \sum_{\langle ij\rangle\in E}f_p^{\langle ij\rangle}(\gb),
\end{align}
where $f^{\langle ij\rangle}_p$ is an individual edge contribution to the total cost function.
The QAOA ansatz only correlates vertices within $p$ steps, so each clause $f^{\langle ij\rangle}_p$ may be computed on a subgraph $S^{\langle ij\rangle}_p$ that only includes the edges that are incident from vertices at a distance $p$ from the vertices $i$ or $j$.
The expectation value thus can be calculated as a sum over $M$ independent subgraphs, as sketched in \mbox{Fig.~\ref{fig:tree_subgraph}A-B}.

Due to this locality, QAOA cannot take advantage of graph structures that are greater than $2p$ steps away~\cite{farhi2014quantum}. Additionally, QAOA cannot distinguish between graphs with cycles of size $>2p+1$, which suggests that worst-case graphs have no small cycles. This intuition was made concrete in~\cite{Wurtz_guarantee}, which proved that worst-case 3 regular graphs with the smallest approximation ratio for $p=1$ and $2$ are bipartite with no cycles $\leq 3$ or $5$, respectively. A bipartite graph has a MaxCut value of $M$, cutting every edge. The only subgraphs of a graph with no small cycles are the tree subgraph, which has no cycles (see Fig.~\ref{fig:tree_subgraph}). The optimized value of a worst-case graph $\mathcal G_*$ is then

\begin{equation}
F_p^{\mathcal G_*}(\gb) = \sum_{\langle ij\rangle \in \mathcal G_*} f^{ij}_p(\gb) = M f_p^\text{tree}(\gamma,\beta).
\end{equation}

Thus, the approximation ratio of a worst-case graph is simply the expectation value of the tree subgraph $f_p^\text{tree}(\gamma,\beta)$. The expectation value of the tree subgraph evaluated at optimal parameters $\{\gamma,\beta\}_\text{tree}$ serves as the performance guarantee. These worst-case graphs are exponentially rare but can be found constructively \cite{cage_survey}.

Additionally, it was proven that any graph evaluated at angles optimum to the tree subgraph have an approximation ratio larger than the guarantee, for $p=1$ and $2$. These facts naturally lead to two conjectures for larger values of $p$ \cite{Wurtz_guarantee}

\quad

\quad
\\
\textbf{Large Loop conjecture:} The worst-case graphs for fixed p are bipartite and have no cycles less than 2p + 2.

\quad
\\
\textbf{Fixed Angle conjecture:} Any graph evaluated at fixed angles optimal to the tree subgraph will have an approximation ratio larger than the guarantee.

\quad

These conjectures are proven \texttt{True} in \cite{Wurtz_guarantee} for $p=1$ and $2$ for 3 regular graphs. These fixed angles act as ``universally good" parameters with good, but not maximum, performance on any 3 regular graph.

This work provides heuristic evidence of these conjectures for larger $p<12$. First, as outlined in section \ref{sec:methods}, we compute expectation values of the tree subgraph $f_p^\text{tree}(\gamma,\beta)$ for $p<12$. Due to the doubly exponential complexity of simulation (the $p=11$ tree subgraph has 8190 vertices, each corresponding to a qubit), we use tensor methods \cite{markov2008simulating} for efficient classical simulation.

Next, we optimize the $2p$ variational parameters via a modified multistart gradient ascent algorithm \cite{gradient_overview} to find optimal angles for worst-case graphs $\{\gamma,\beta\}_\text{tree}$, which serve as fixed angles to evaluate on any graph of the same regularity. The optimal expectation values of the tree subgraph serve as a performance guarantee assuming the large loop conjecture.

Finally, as outlined in section \ref{sec:results}, we provide a heuristic proof of the fixed angle conjecture by evaluating the approximation ratio on all 3 regular graphs with $\leq 16$ vertices. Further, we index the merits of these pre-computed, fixed angles to speed up and improve the variational optimization step of QAOA. We conclude in section \ref{sec:results} with some of the implications of these fixed angle conjectures, including bounds for quantum advantage on regular graphs.

\section{Methodology} \label{sec:methods}

\subsection{Simulations}
A necessary step of QAOA is evaluating the expectation value of the cost function.
One option is to compute the value on quantum hardware, by
sampling from the ansatz state $|\gamma,\beta\rangle$ and calculating the expectation value statistically.
While in principle this is the only method that will work for arbitrary circuit parameters, it is possible to simulate expectation values on a classical computer for some circuits even of extensive size. 

One simple alternate method is state-vector evolution, which requires storing the $2^N$ values of the wavefunction in memory, and evolving via sparse matrix exponentiation methods. However, this method is infeasible for large system sizes of $\gtrsim 20$ qubits, due to the exponentially scaling memory requirements.

For this work, we use the classical simulator \texttt{QTensor} \cite{qtensor, lykov2021large} which is based on tensor network contraction and allows for simulation of a much larger number of qubits for a limited set of wavefunctions. 
Instead of storing the full state vector of the system and evolving it by applying matrix transformations, the state is represented by a tensor network, and the gates with tensors that have input and output indices. The tensor network constructed in this way can then be contracted in an efficient manner to compute expectation values.

While there may exist multiple approaches for determining the best way to contract a tensor network, we use a contraction approach called bucket elimination \cite{detcher2013bucket}, which contracts indices of the tensor expression sequentially. At each step, we choose some index $j$ from the tensor expression and then sum over a product of tensors that have $j$ in their index. The size of the intermediary tensor obtained as a result of this operation is very sensitive to the order in which indices are contracted.
To find a good contraction ordering we use a line graph \footnote{\url{https://en.wikipedia.org/wiki/Line_graph}} of the tensor network.
A tree decomposition \footnote{\url{https://en.wikipedia.org/wiki/Tree_decomposition}} of the line graph corresponds to a contraction path that guarantees that the number of indices in the largest intermediary tensor will be equal to the width of the tree decomposition \cite{markov2008simulating}.
Figure \ref{fig:tree_subgraph}C shows the line graph of the tensor network that corresponds to calculating edge contribution $f^{\langle ij\rangle}_p$ for subgraph shown on Fig. \ref{fig:tree_subgraph}B. In this way, it is possible to simulate QAOA to reasonable depth on hundreds or thousands of qubits. More details of \texttt{QTensor} and tensor networks are in \cite{shutski2019adaptive, lykov2021large, Gray_Cotengra,lykov2021importance}. We observe that the tree subgraph has a particularly simple and symmetric tensor structure, which allows for efficient contraction for depths $p\sim 11$ which would not be possible for more general and generic subgraphs.

\subsection{Angle optimization}

A necessary step of QAOA is optimizing the expectation value of the cost function with respect to the variational parameters $\gamma,\beta$.
One common optimization method is gradient ascent, which moves the parameter along the direction of the steepest gradient. To compute gradients, we use automatic differentiation provided by PyTorch and \texttt{QTensor}. To speed optimization, we use the modified gradient descent algorithm \texttt{RMSprop} \cite{gradient_overview}, an established machine learning algorithm.

The optimization surface has multiple local maxima, which provide a challenge to gradient descent algorithms. For this reason, we choose two initialization routines. The first routine is that of Refs.~\cite{zhou2020,wurtz2021counterdiabaticity}, which initializes parameters in a counterdiabatic configuration. Starting with $p=2$, larger $p+1$ are interpolated by deriving the underlying continuous counterdiabatic schedule and matching angles. The gradient ascent initialized from these angles finds a local optimum with smooth angles, although there is no guarantee that these smooth angles are a global optimum.

To verify that the smooth angles are global optimum, we also implemented a multistart routine \cite{Shaydulin_Multistart}, which executes parallel optimization initialized from 1000 random points in parameter space. With high probability, one of the random initial points is in the same basin of attraction as the global optimum, and so the routine returns the optimal angles. We find for $p\leq 6$ that the global optima returned by this method returns the same value as the smooth angles within numerical precision, indicating that the angles for $p\leq 6$ are global optima. For $p>6$, multistart always returned lower results than the counterdiabatic-initialized angles.

\section{Results}\label{sec:results}

Here, we present the fixed angles $\{\gamma,\beta\}_\text{tree}$ as well as the heuristic evidence of the fixed angle conjecture. Using the methods of Sec.~\ref{sec:methods}, we compute optimal parameters for the tree subgraph $\{\gamma,\beta\}_\text{tree}$ via a loop between the \texttt{QTensor} tensor simulator and the classical optimization routine.

Optimized angles for 3 regular graphs and $p<12$ are shown in Table \ref{table:3r_angles}. We provide this data in a machine-readable \texttt{JSON} format for regular graphs of degrees 3 to 5 as a supplemental to this paper \cite{fixed-angle-QAOA-github}. These angles are also available to use in the \texttt{QTensor} package as \texttt{qtensor.tools.BETHE\_QAOA\_VALUES}.

\subsection{Guaranteed performance}\label{sec:guaranteed_perf}

\begin{figure} 
\centering
\includegraphics[width=\linewidth]{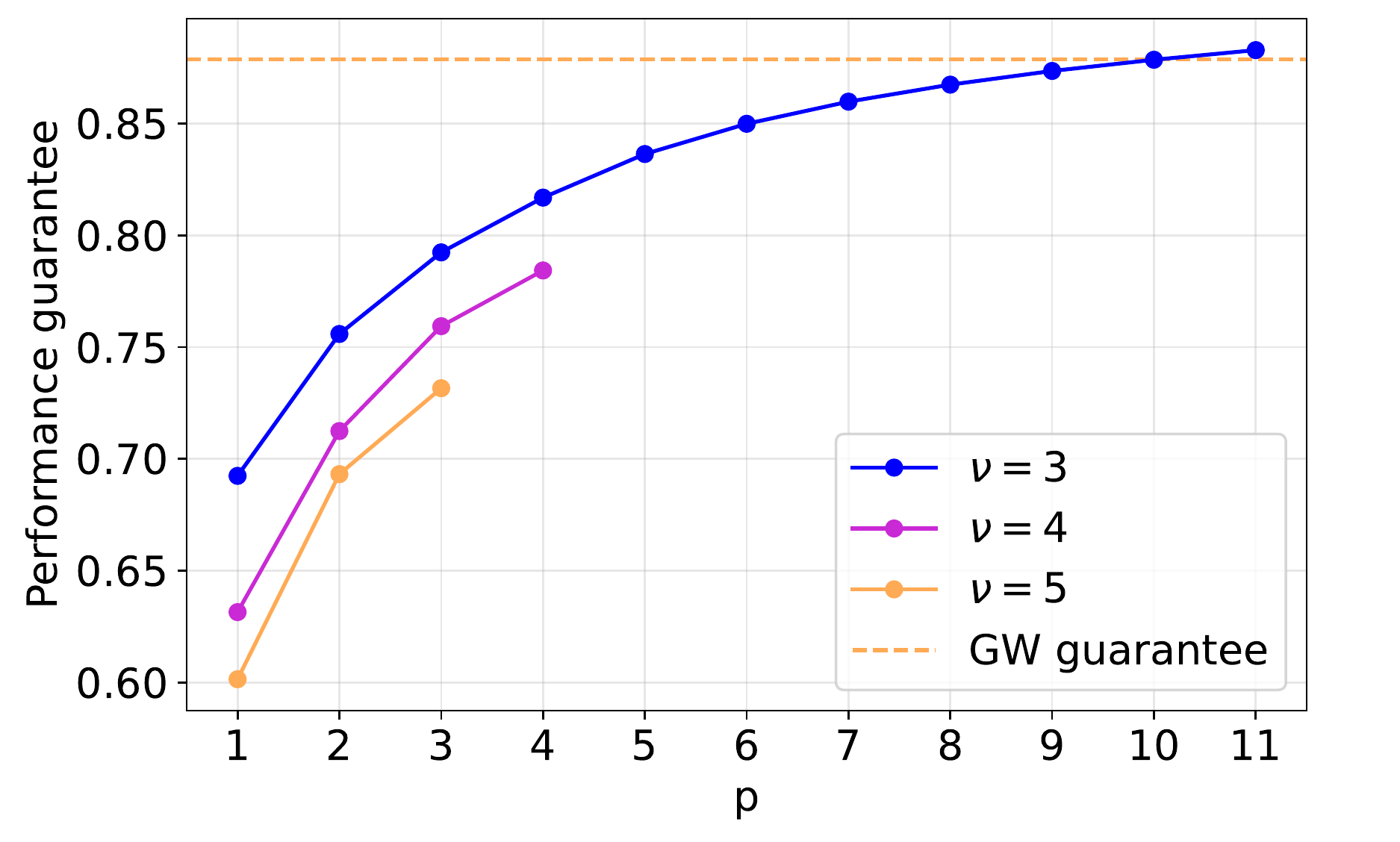}
\caption{The performance guarantee for $\nu$-regular graphs as a function of $p$, assuming the large loop conjecture where worst-case graphs are bipartite and have no small cycles. The guarantee appears to approach unity as $1/\sqrt{p}$, and surpasses the guarantee of Goemans Williamson (orange dashed) for $p\geq 11$. The performance guarantee for larger connectivity is smaller and goes as $1/\sqrt{\nu}$ for fixed $p$, consistent with a mean field picture.
}
\label{fig:ar_vs_conn}
\end{figure}

\begin{figure*}
\centering
\includegraphics[width=\linewidth]{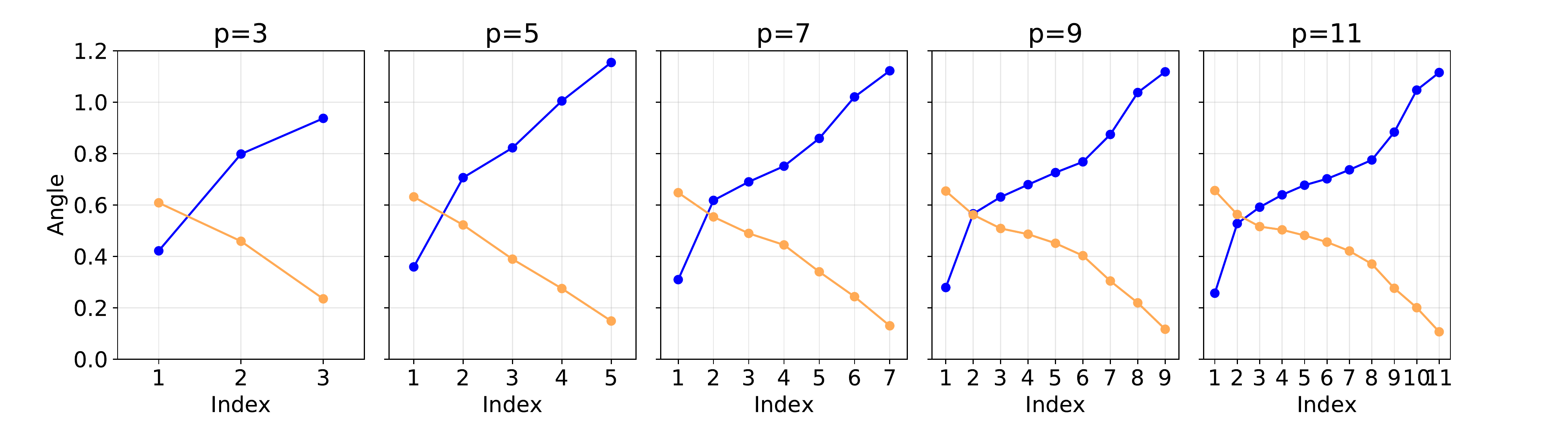}
\label{fig:gamma_beta}
\end{figure*}

\begin{table*}[]
\begin{tabular}{|c|c|cccccccccccc|}
\hline
\multicolumn{2}{|l|}{} &       &1 &2 &3 &4 &5 &6 &7 &8 &9 &10 &11 \\ \hline
\multirow{2}{*}{\;$p=1$\;}     &  \multirow{2}{*}{$C_{1}\geq0.6925$}&$\gamma$& \; 0.616\; & & & & & & & & & & \\ 
                       &  & $\beta$  & \; 0.393\; & & & & & & & & & & \\ \hline
\multirow{2}{*}{\;$p=2$\;}     &  \multirow{2}{*}{$C_{2}\geq0.7559$}&$\gamma$& \; 0.488\; & \; 0.898\; & & & & & & & & & \\ 
                       &  & $\beta$  & \; 0.555\; & \; 0.293\; & & & & & & & & & \\ \hline
\multirow{2}{*}{\;$p=3$\;}     &  \multirow{2}{*}{$C_{3}\geq0.7924$}&$\gamma$& \; 0.422\; & \; 0.798\; & \; 0.937\; & & & & & & & & \\ 
                       &  & $\beta$  & \; 0.609\; & \; 0.459\; & \; 0.235\; & & & & & & & & \\ \hline
\multirow{2}{*}{\;$p=4$\;}     &  \multirow{2}{*}{$C_{4}\geq0.8169$}&$\gamma$& \; 0.409\; & \; 0.781\; & \; 0.988\; & \; 1.156\; & & & & & & & \\ 
                       &  & $\beta$  & \; 0.600\; & \; 0.434\; & \; 0.297\; & \; 0.159\; & & & & & & & \\ \hline
\multirow{2}{*}{\;$p=5$\;}     &  \multirow{2}{*}{$C_{5}\geq0.8364$}&$\gamma$& \; 0.360\; & \; 0.707\; & \; 0.823\; & \; 1.005\; & \; 1.154\; & & & & & & \\ 
                       &  & $\beta$  & \; 0.632\; & \; 0.523\; & \; 0.390\; & \; 0.275\; & \; 0.149\; & & & & & & \\ \hline
\multirow{2}{*}{\;$p=6$\;}     &  \multirow{2}{*}{$C_{6}\geq0.8499$}&$\gamma$& \; 0.331\; & \; 0.645\; & \; 0.731\; & \; 0.837\; & \; 1.009\; & \; 1.126\; & & & & & \\ 
                       &  & $\beta$  & \; 0.636\; & \; 0.534\; & \; 0.463\; & \; 0.360\; & \; 0.259\; & \; 0.139\; & & & & & \\ \hline
\multirow{2}{*}{\;$p=7$\;}     &  \multirow{2}{*}{$C_{7}\geq0.8598$}&$\gamma$& \; 0.310\; & \; 0.618\; & \; 0.690\; & \; 0.751\; & \; 0.859\; & \; 1.020\; & \; 1.122\; & & & & \\ 
                       &  & $\beta$  & \; 0.648\; & \; 0.554\; & \; 0.490\; & \; 0.445\; & \; 0.341\; & \; 0.244\; & \; 0.131\; & & & & \\ \hline
\multirow{2}{*}{\;$p=8$\;}     &  \multirow{2}{*}{$C_{8}\geq0.8674$}&$\gamma$& \; 0.295\; & \; 0.587\; & \; 0.654\; & \; 0.708\; & \; 0.765\; & \; 0.864\; & \; 1.026\; & \; 1.116\; & & & \\ 
                       &  & $\beta$  & \; 0.649\; & \; 0.555\; & \; 0.500\; & \; 0.469\; & \; 0.420\; & \; 0.319\; & \; 0.231\; & \; 0.123\; & & & \\ \hline
\multirow{2}{*}{\;$p=9$\;}     &  \multirow{2}{*}{$C_{9}\geq0.8735$}&$\gamma$& \; 0.279\; & \; 0.566\; & \; 0.631\; & \; 0.679\; & \; 0.726\; & \; 0.768\; & \; 0.875\; & \; 1.037\; & \; 1.118\; & & \\ 
                       &  & $\beta$  & \; 0.654\; & \; 0.562\; & \; 0.509\; & \; 0.487\; & \; 0.451\; & \; 0.403\; & \; 0.305\; & \; 0.220\; & \; 0.117\; & & \\ \hline
\multirow{2}{*}{\;$p=10$\;}     &  \multirow{2}{*}{$C_{10}\geq0.8785$}&$\gamma$& \; 0.267\; & \; 0.545\; & \; 0.610\; & \; 0.656\; & \; 0.696\; & \; 0.729\; & \; 0.774\; & \; 0.882\; & \; 1.044\; & \; 1.115\; & \\ 
                       &  & $\beta$  & \; 0.656\; & \; 0.563\; & \; 0.514\; & \; 0.496\; & \; 0.469\; & \; 0.436\; & \; 0.388\; & \; 0.291\; & \; 0.211\; & \; 0.112\; & \\ \hline
\multirow{2}{*}{\;$p=11$\;}     &  \multirow{2}{*}{$C_{11}\geq0.8828$}&$\gamma$& \; 0.257\; & \; 0.528\; & \; 0.592\; & \; 0.640\; & \; 0.677\; & \; 0.702\; & \; 0.737\; & \; 0.775\; & \; 0.884\; & \; 1.047\; & \; 1.115\; \\ 
                       &  & $\beta$  & \; 0.656\; & \; 0.563\; & \; 0.516\; & \; 0.504\; & \; 0.482\; & \; 0.456\; & \; 0.421\; & \; 0.371\; & \; 0.276\; & \; 0.201\; & \; 0.107\; \\ \hline
\end{tabular}
        \caption{
        The fixed 3 regular $\gammabeta$ QAOA angles optimal to the tree subgraph. Under the fixed angle conjecture, 3 regular graphs evaluated at these angles will have an approximation ratio larger then the performance guarantee of column 2. Angles are normalized to be consistent with Eq.~\eqref{eq:QAOA_ansatz}. We find this conjecture to be \texttt{True} on all graphs with $\leq 16$ vertices, as shown in Fig.~\ref{fig:heuristic_AR}. A \texttt{JSON} file with these angles, plus fixed angles for regular graphs of degree 4 and 5, is provided as a supplemental \cite{fixed-angle-QAOA-github}.
        }
        \label{table:3r_angles}
\end{table*}
    
    Under the large loop conjecture, the expectation value of the tree subgraph serves as a QAOA performance guarantee for any regular graph of the same degree. 
    By evaluating the tree subgraph at these optimal angles, the performance guarantee for 3 regular graphs is shown in Table \ref{table:3r_angles}, and plotted in Fig.~\ref{fig:ar_vs_conn} for regular graphs of degree 3 to 5. For $p\geq 11$, the performance guarantee is $C_{11}\geq 0.8828$. This threshold is important, as it is larger than the performance guarantee of the best general purpose algorithm of Goemans and Williamson (GW)\cite{gw-algo} which has a performance guarantee of 0.8786. Having larger performance guarantees than competing classical algorithms is one indicator of quantum advantage \cite{Bravyi2018advantage}; however, there is no advantage here. The GW algorithm is general-purpose and works on any graph of any connectivity and edge weights, while these performance guarantees only work for 3 regular graphs and fixed edge weights. There exist better special-purpose implementations of GW for 3 regular graphs with larger performance guarantees. For example, Ref. \cite{HALPERIN2004169} outlines a semidefinite programming solver similar to GW specialized to 3 regular graphs, with an approximation ratio $C>0.9326$. These fixed angles do not achieve advantage over the ``best" classical algorithms. However, this still an important step towards quantum advantage for QAOA.

    There also exist powerful heuristic solvers \cite{gurobi, MQLib}, which do not provide any guaranteed performance, but are designed to perform as well on typical graphs.
    For example, we use the Gurobi algorithm~\cite{gurobi} to exactly solve 3-regular graphs of sizes up to 256, which corresponds to approximation ratio of 1.
    Nevertheless, there is still a chance that for a particular graph a heuristic solver will perform extremely poorly. Since the fixed angle QAOA provides \emph{guaranteed} performance, we focus on comparing it with classical algorithms that also provide such guarantees.
    
    We observe that the performance guarantee as a function of $p$ appears to grow slower as $p$ increases. However, for large $p$, the performance guarantee must converge to a finite value $\leq 0.9351$, due to indistinguishability \cite{farhi2020quantum2}. For any girth $g$, there exist 3r graphs which have a cut fraction of at most $0.9351$ \cite{Kardo2012}. Such graphs have high girth, and so is constructed only of the tree subgraph. The approximation ratio is bounded from above by 1, and so the value $f_p\leq 0.9351$. This suggests that guaranteed advantage may never occur, as the best classical algorithms have a guarantee of $0.9326$.

    We also find that the performance guarantees for larger degree are smaller (Fig.~\ref{fig:ar_vs_conn}), and scale as $1/\sqrt{\nu}$ consistent with a mean-field picture. This result suggests that, contrary to common lore \cite{farhi2020quantum}, low connectivity graphs may have better performance than high connectivity graphs, at least in terms of performance guarantees.
    
    \begin{figure} 
        \centering
        \includegraphics[width=\linewidth]{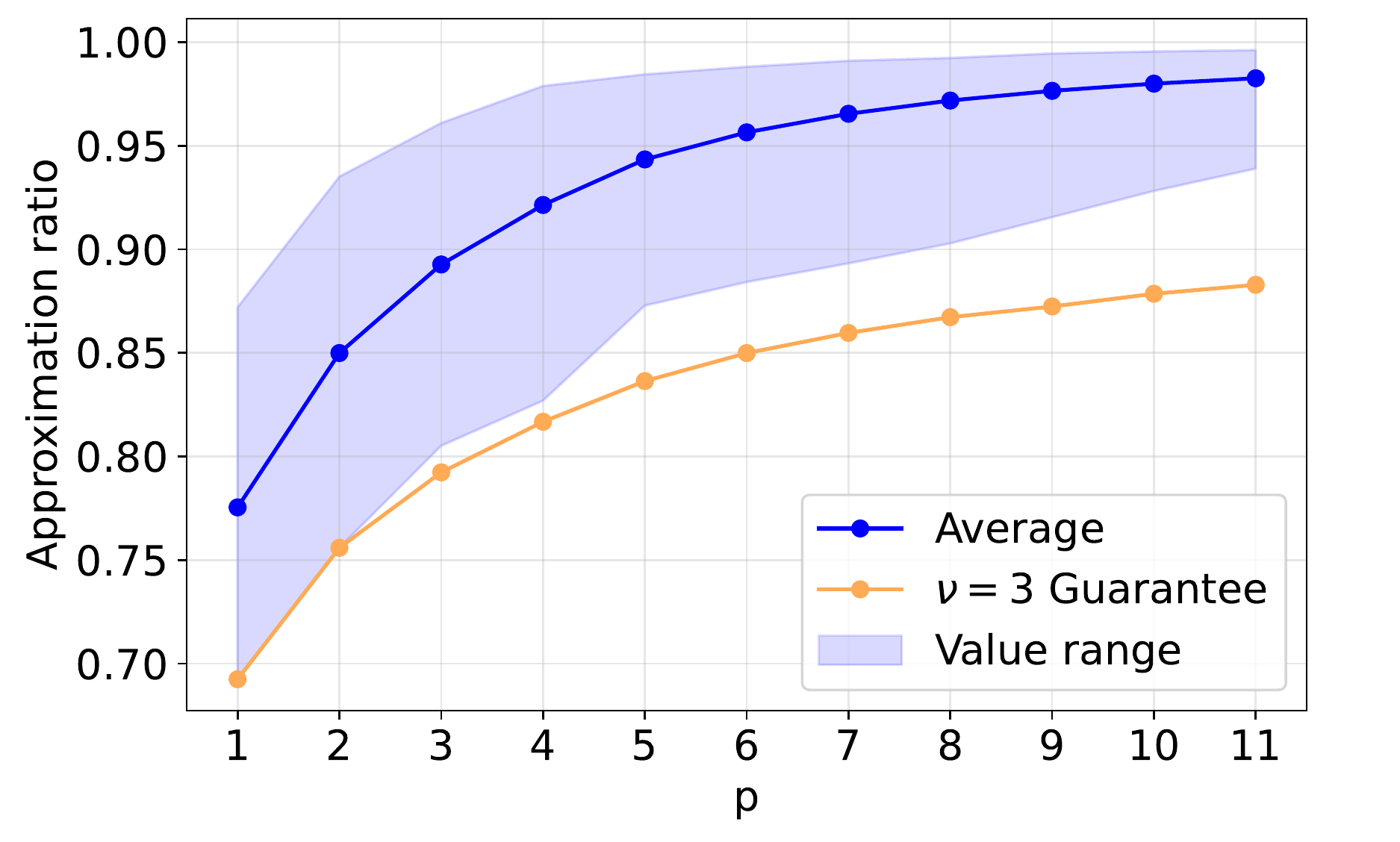}
        \caption{Approximation ratio for an ensemble of all 4681 3-regular graphs with $\leq 16$ vertices. For each graph and depth, QAOA is evaluated at the angles shown in Table~\ref{table:3r_angles}, then divided by the optimal MaxCut value for the graph.
       The shaded region shows the absolute range between the worst and best approximation ratios of the ensemble. The yellow line shows the guarantee for any 3-regular graphs. This plot is numerical evidence of the fixed angle conjecture, which states that the approximation ratio for any 3-regular graph evaluated at fixed angles is larger than the guarantee.
        }
        \label{fig:heuristic_AR}
    \end{figure}
    
    \begin{figure} 
        \centering
        \includegraphics[width=\linewidth]{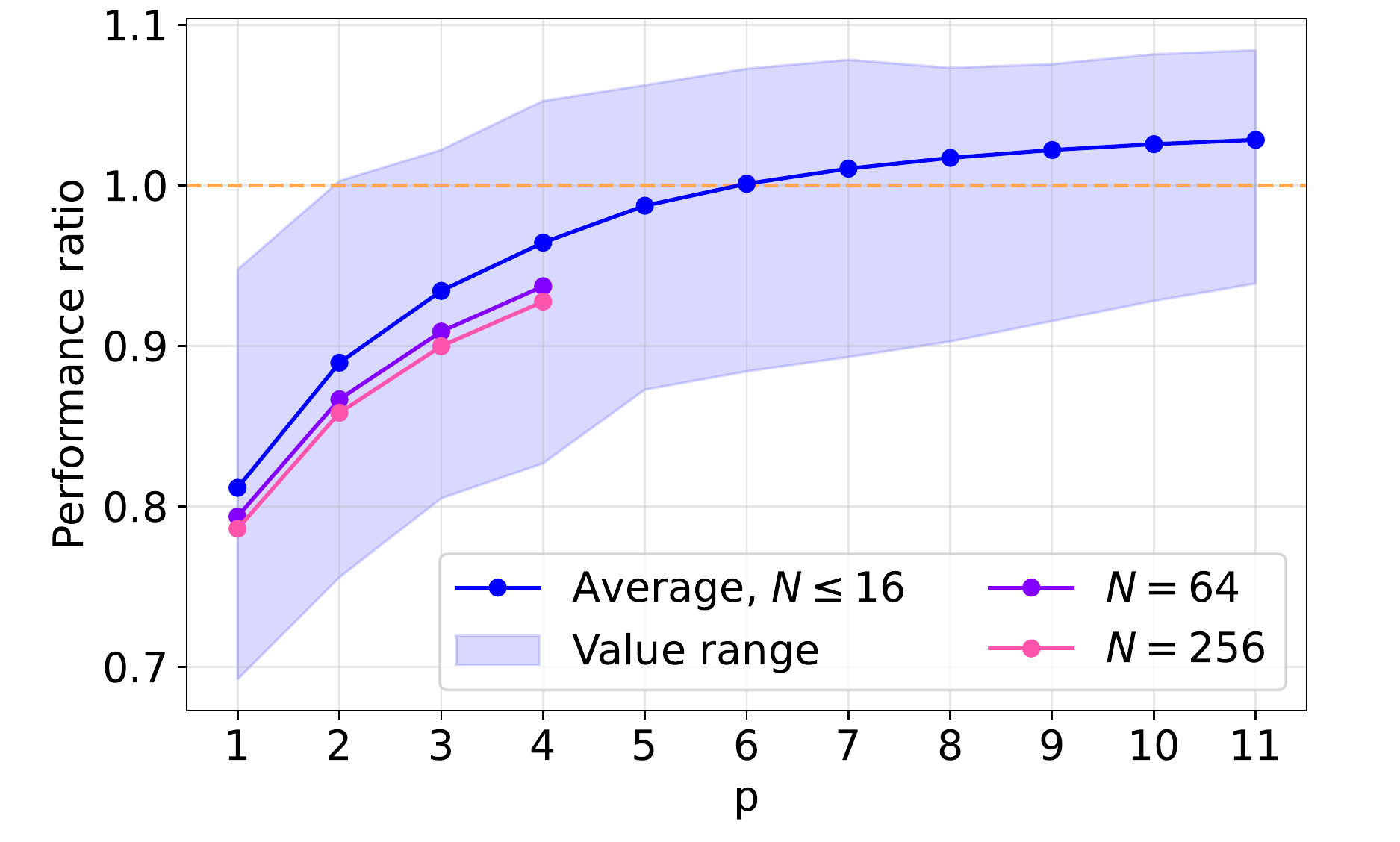}
        \caption{
        Performance ratio as defined in Eq.~\eqref{eq:perf_ratio} vs.~the classical GW algorithm, for the ensemble of all 4681 3 regular graphs with $\leq 16$ vertices. For each graph and depth, QAOA is evaluated at the angles shown in Table~\ref{table:3r_angles}, then divided by the average approximation ratio of solutions given by GW for the graph.
        The shaded region shows the absolute range between the worst and best performance ratios of the ensemble. The dashed yellow line corresponds to parity between GW and QAOA.
        For $p\geq6$, QAOA with fixed angles has an average case advantage over the classical GW algorithm. Purple and pink plot the average performance ratio over an ensemble of 32 graphs with $N$ vertices, indicating a reduced but increasing with $p$ performance as a function of $N$.
        }
        \label{fig:heuristic_PR}
    \end{figure}
    
    \subsection{Heuristic performance}\label{sec:heuristic}
    
    In Sec.~\ref{sec:guaranteed_perf}, we provide strict performance guarantees for regular graphs, which for $p>2$ only assume the large loop conjecture. However, QAOA is usually considered a heuristic algorithm, where performance is surveyed over an ensemble of graphs. While the guaranteed performance of QAOA may be small, the worst-case bipartite graphs with no small cycles are extremely atypical. The approximation ratio of any particular typical graph may be much larger than the guarantee.
    
    Here, we provide heuristic evidence for the fixed angle conjecture, which posits that any graph evaluated at these fixed angles have an approximation ratio larger than the guarantee. Further, we will show that typical performance does not saturate performance guarantees, and is competitive with the GW algorithm even at fixed angles.
    
    As an ensemble, we choose all 3 regular graphs with $\leq 16$ vertices. There are 4681 such graphs \cite{Meringer1999}\footnote{ We use \url{https://hog.grinvin.org/Cubic} for enumeration.}. For each graph, we evaluate the expectation value $F_p^{\mathcal G}(\gamma,\beta)$ at the fixed angles of Table \ref{table:3r_angles}. Note that because the optimization step is bypassed, the QAOA execution is very fast ($<1$sec), as it requires only one query to the quantum simulator. Using an exhaustive search we find the exact MaxCut value $C_\text{max}^{\mathcal G}$ to compute the approximation ratio $C_p^{\mathcal G}$. Data for this ensemble of graphs is shown in Fig.~\ref{fig:heuristic_AR}. It is clear that the minimum approximation ratio, representing the worst-case over the ensemble, is always greater than the performance guarantee. This is heuristic evidence of the fixed angle conjecture: there are no graphs with $\leq 16$ vertices evaluated at these fixed angles which are below the worst-case guarantee. To prove the fixed angle conjecture \texttt{False}, one needs simply to provide a 3 regular graph whose approximation ratio is less than the guarantee.
    
    Although rare, the worst-case in the ensemble may still saturate the performance guarantee if the graph is bipartite with no cycles $\leq 2p+1$, as can be seen for $p=1$ and $2$. The smallest worst-case graph is a ``cage" \footnote{\url{https://mathworld.wolfram.com/CageGraph.html}}. The smallest worst-case $p=2$ graph has 14 vertices while the smallest worst-case graph for $p=3$ has 30 vertices. Such worst-case graphs must be at least exponentially large in $p$ to fit the exponentially large tree subgraph; a $p=11$ graph must have at least 8190 vertices, and more to satisfy the condition of having no small cycles and bipartite. A minimal cage graph is a Moore graph \cite{Godsil2001}.

    While quantum advantage in terms of performance guarantees occurs for $p\geq 11$, it is also interesting to check performance in comparison to competing classical algorithms. If the quality of the solution returned by a quantum algorithm is larger than that returned by the competing classical algorithm for a particular graph, then the quantum algorithm has advantage over the particular classical algorithm for the particular graph.
    This condition is parameterized by the performance ratio
    
    \begin{equation}
        B_p^{\mathcal G}(\gamma,\beta) \;=\;   F_p^{\mathcal G}(\gamma,\beta)\;\;/\;\;\langle C_\text{cl}^{\mathcal G}\rangle
        \label{eq:perf_ratio}
    \end{equation}
    where $\langle C_\text{cl}^{\mathcal G}\rangle$ is the average number of cut edges returned by the classical algorithm, if the algorithm is non-deterministic. If $B_p^{\mathcal G}>1$, the quantum algorithm has advantage for the particular graph $\mathcal G$, as it will return solutions which are better on average than the classical algorithm. If $\langle B_p^{\mathcal G}\rangle >1$, where $\langle*\rangle$ indicates average over some graph ensemble $\{\mathcal G\}$, then the algorithm has average case advantage over the graph ensemble.
    The performance ratio is bounded from below by the performance guarantee of the quantum algorithm, and bounded from above by the inverse performance guarantee of the classical algorithm.
    
    We evaluate the performance ratio over the same ensemble of all 3 regular graphs with $\leq 16$ vertices in Fig.~\ref{fig:heuristic_PR}. The QAOA expectation value is evaluated at the fixed angles $\{\gamma,\beta\}_\text{tree}$, and the classical expectation value is evaluated using 100 queries of the Goemans Williamson algorithm \footnote{We use the python package \texttt{cvxgraphalgs} to find solutions \texttt{cvxgraphalgs.algorithms.goemans\_williamson\_weighted}. \url{https://pypi.org/project/cvxgraphalgs/}}. For all $p$ evaluated, there exist graphs in the ensemble which do not have advantage (lower edge of the shaded region). However, for $p\geq 2$ there exist graphs in the ensemble which do have advantage (upper edge of the shaded region); the only graph with advantage for $p=2$ is the 10 vertex Peterson graph, which is the smallest graph of girth 5. For $p\geq 6$ the QAOA at fixed angles has average case advantage over the ensemble of all graphs with $\leq 16$ vertices.
    
    While Fig.~\ref{fig:heuristic_PR} suggest average case advantage for $p\geq 6$, there is the caveat that it may be a result of particular choice of small graphs in the ensemble. To strengthen the evidence of fixed angle QAOA outperforming the classical GW algorithm, we next evaluate the performance on larger graphs.
    
    \subsection{Heuristic performance for larger N}\label{sec:heuristics_largeN}

    While these heuristic results are strong evidence of the fixed angle conjecture and an optimization-free QAOA, the argument is weakened by the fact that the ensemble is only over small graphs. Here, we strengthen the argument by evaluating the approximation ratio at fixed angles over an ensemble of graphs of increasing sizes $N\in \{\,8,\dots,256\,\}$.
    While state-vector evolution is infeasible for such exponentially large Hilbert spaces, tensor network computation of expectation values is still feasible, at least for $p\leq 4$. The size of subgraphs are exponential with $p$ and so the calculations are unfeasible for larger $p$.
    Similarly, while brute enumeration of all solutions is infeasible to compute the exact MaxCut value, we use the industry-standard solver \texttt{GUROBI} \cite{gurobi} to find MaxCut values. The solver still requires exponential time, but with a reduced prefactor, enabling computation of exact MaxCut values for a $256$ vertex graph in $\lesssim 20$sec.
    
    \newcommand{\numNgraphs}{{32}}

\begin{figure} 
        \centering
        \includegraphics[width=\linewidth]{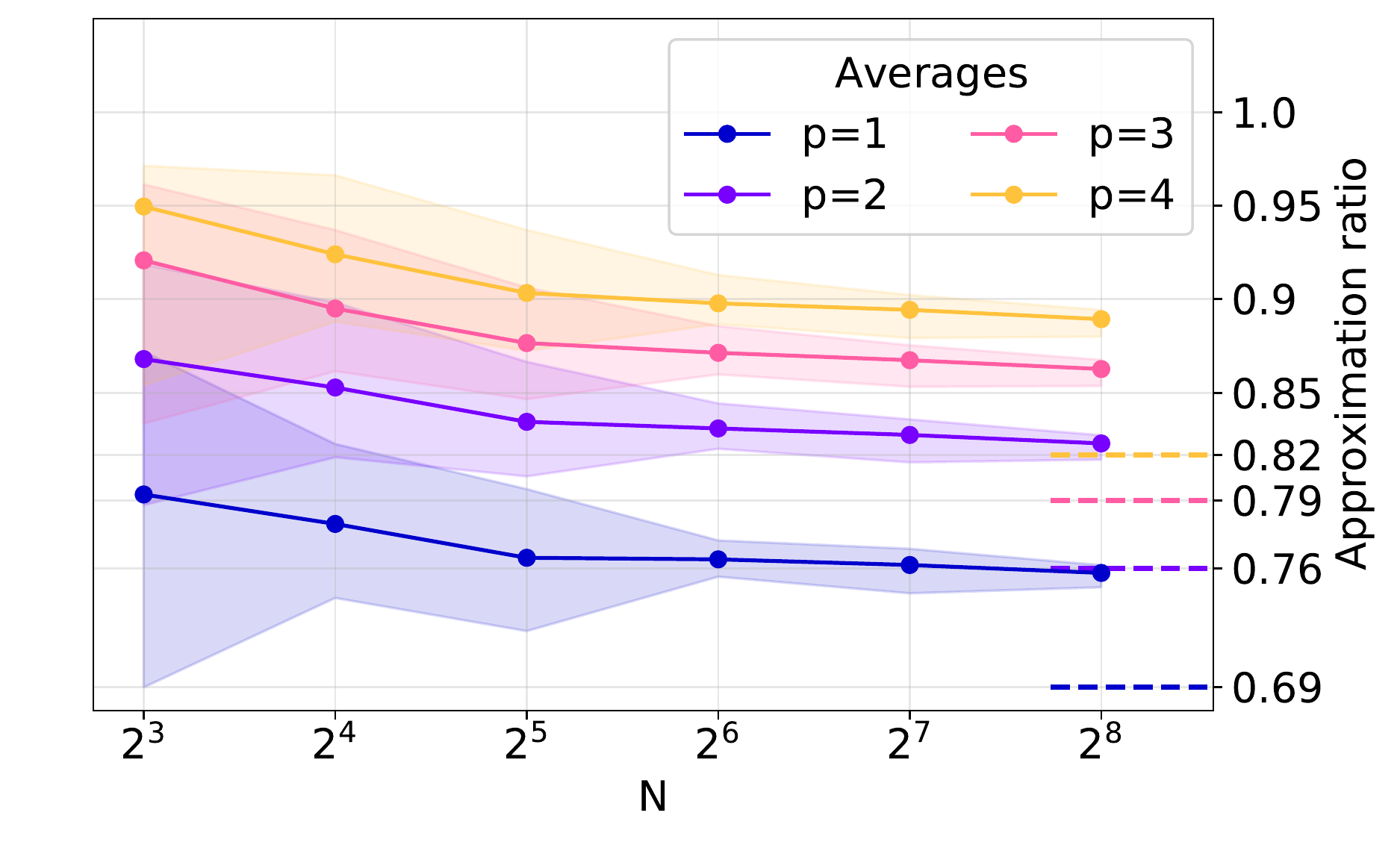}
        \caption{Approximation ratio as a function of $N$ and $p$. Each point is the average over an ensemble of {\numNgraphs } 3-regular graphs of $N$ vertices. The shaded region shows the absolute range between the worst and best approximation ratios of the ensemble. Average approximation ratios appear to slowly converge to some $N$-independent value, as expected, and no graph violates the fixed angle conjecture (dashed).}
        \label{fig:heuristic_AR_funcn_N}
    \end{figure}

    Results for the approximation ratio of this ensemble are shown in Fig.~\ref{fig:heuristic_AR_funcn_N}. For each size $N$, we choose \numNgraphs~graphs as a random subset of all 3 regular graphs with $N$ vertices. For each graph, we evaluate the expectation value $F_p^{\mathcal G}(\gamma,\beta)$ at the fixed angles of Table \ref{table:3r_angles} using \texttt{QTensor}.  For smaller $N$, the approximation ratio is larger and, as N grows, converges to some size-independent typical value. For all $p$ and $N$ chosen, the approximation ratio of each graph in the ensemble is larger than the guarantee, which further strengthens the fixed angle conjecture. Similarly, results for the performance ratio of this ensemble are shown in Fig.~\ref{fig:heuristic_PR}. For larger graphs, the performance decreases, but may converge to some size-independent value. This is consistent with the fact that the GW algorithm also has consistent performance across a range of graph sizes. Due to this decrease in performance, average-case advantage for large graphs may occur for slightly larger $p$, although the crossover is yet to be determined.
    
    The decrease in approximation ratio as function of $p$ is expected. For $p=1$, the QAOA is only sensitive to graph structure within 3 steps, and ultimately only ``sees" cycles of size 3 in the graph. The fraction of edges which participates in a cycle of size 3 is asymptotically zero for $N\to\infty$ \cite{WORMALD1981} and so for large $N$ the only subgraphs of a typical graph are the tree subgraph, and is close to worst-case. Such a typical graph would saturate the performance guarantee if it was bipartite; however, bipartite graphs are atypical \footnote{Bipartite graphs: \url{http://oeis.org/A005142}\\ All graphs: \url{https://oeis.org/A001349}}. Thus, the average approximation ratio for large $N$ is the performance guarantee, divided by the average fraction of MaxCut edges cut, which we observe to converge to 0.91 as $N$ grows \cite{Dembo_2017}.
    A similar argument holds for larger $p$, except with a sensitivity to larger cycles. The fraction of edges that participate
    in a cycle of a fixed size approaches zero with $N\rightarrow\infty$, with logarithmically slower convergence for large cycles \cite{WORMALD1981}. Consequentially, the average approximation ratios also converge to asymptotic values, although logarithmically slower for larger $p$.
    
    \subsection{Fixed angles vs.~global optima}
    
    While the fixed angle conjecture states that fixed angles will have good performance, as the approximation ratio must be above a reasonably large guarantee, it is interesting to compare fixed angles to global maxima. How close are these fixed angles to optimal angles?
    
    To investigate the relative performance of these fixed angles vs. global maxima, we compute their global optima for the ensemble of all 3 regular graphs with $N\leq 16$ and $p\leq 2$. These optima are found via the same multistart gradient ascent procedure of section \ref{sec:methods}.

    Heuristic results comparing the approximation ratios of the global maxima vs fixed angles are shown in Table \ref{tab:parameter_transfer}. On average (rows 1-3), the difference in the approximation ratio between global optimum and that evaluated at fixed angles is less than 0.003, at least over all small graphs. These values indicate that the fixed angles are extremely close to optimal. This suggests that using these fixed angles can allow QAOA to bypass the variational optimization step by simply evaluating a regular graph at these pre-computed angles, and get an approximation ratio which is almost the same as the global optimum.
    
    That these pre-computed angles are very close to optimal is not a priori obvious. However, this phenomenon has been heuristically observed before as the transfer of parameters \cite{brandao2018fixed,streif2019training,galda2021transferability}, which observes that optimal parameters for one graph are good for other graphs in the same class. These fixed angle results are potentially an underlying explanation for this observation: optimal angles for an ensemble of graphs may concentrate in a small region of parameter space around these fixed angles.
    
    This conjecture about parameter concentration can be heuristically checked by checking if the fixed angles are close to global optima. Note that this task is nontrivial: both optimal angles, and fixed angles, are degenerate (see for example, \cite{Wurtz_guarantee} Table 1), which excludes evaluating closeness by Euclidean distance between particular optima. Instead, we use the fixed angles as a warm start for the gradient ascent variational optimization procedure. If the fixed angles are in the same basin of attraction as a global optimum, the optimization returns the global optima and the fixed angles are ``close". The probability that this occurs is indexed in Table \ref{tab:parameter_transfer}. For a vast majority of graphs in the small ensemble, the warm start gradient ascent optimization finds the global maximum (row 4). Additionally, the gradient ascent procedure changes the parameters very little. As shown in row 5, the euclidean distance between the optimized angles and initial point is usually very small (in units of $\pi)$. These results suggest that using these fixed angles as an initial guess may speed up the optimization, and that optimization does not give a lot of improvement.
    
    The fact that fixed angles are so close to global optima, at least for small $p$ and graphs, suggest that we can completely bypass the variational optimization step of QAOA for MaxCut on regular graphs. Instead of having a variational optimization step, one may pre-compute a fixed circuit with the objective function set by the graph, and fixed angles from table \ref{table:3r_angles} chosen for a particular connectivity and $p$. In particular, this may yield a significant speed up in computation: optimization requires many repetitions of querying various points in parameter space to compute low-noise expectation values of the objective function. In opposition, using fixed angles may yield close to optimal bitstrings even if the quantum device is only queried a few times \cite{streif2019training}, which may be a factor of 100-1000$\times$ speedup and enable real-time solutions.

\begin{table} 
\begin{tabular}{|c|c|c|}
    \hline
    QAOA depth $p$& \;\;\; $p=1$\;\;\; & \;\;\;$p=2$\;\;\;\\ \hline
    \begin{tabular}[c]{@{}c@{}}Average approximation ratio\\ evaluated at fixed angles\end{tabular}    &  0.7754    &  0.8499\\\hline
\begin{tabular}[c]{@{}c@{}}Average approximation ratio\\ evaluated at optimal angles\end{tabular}    &  0.7764    &  0.8522\\\hline
\begin{tabular}[c]{@{}c@{}}Average difference between\\ fixed and optimal AR\end{tabular}    &  \;\;0.001016\;\;    &  \;\;0.002288\;\;\\\hline
\begin{tabular}[c]{@{}c@{}}Global optima found\\from fixed angles\end{tabular}    &  
   \begin{tabular}[c]{@{}c@{}}\;\;4681 / 4681\;\; \\ (100.00\%)\end{tabular} & 
      \begin{tabular}[c]{@{}c@{}}\;\;4655 / 4681\;\; \\ (99.445\%)\end{tabular}\\\hline
\begin{tabular}[c]{@{}c@{}}Euclidian distance between\\fixed and optimized\end{tabular}    &  0.0175 & 0.0410\\\hline
\end{tabular}
\caption{Comparing the approximation ratio of graphs evaluated at global optima vs.~fixed angles. Ensemble is all 4681 3 regular graphs with $\leq 16$ vertices. Rows 1 and 2 index the average approximation ratio of the ensemble, evaluated at fixed angles (1) vs. globally optimal angles (2). The average improvement of evaluating at optimal angles is, on average, very small, as indexed by row (3). Row (4) indexes the number of graphs for which a gradient ascent initialized at the fixed angles finds some global optima for the graph. Row (5) is the average Euclidean difference between the fixed angle and the gradient ascent optimized angles.  For $p=1$, this warm start finds the optima for every graph, and for $p=2$ finds the optima for an overwhelming majority. These results indicate that these angles are very good for a majority of graphs, and serve as a good initial guess for optimizers.}\label{tab:parameter_transfer}
\end{table}

    \section{Conclusion}
    
    In this work, we provide numerical evidence for the fixed angle conjecture of Ref. \cite{Wurtz_guarantee}, which states that any graph evaluated at fixed angles will have an approximation ratio larger than the guarantee. This conjecture has the interesting implication that, for regular graphs of fixed degree and constant edge weight, there is a set of angles that are ``universally good" for any graph, in that they may not be global optima but still guarantee good performance.
    
    This evidence was provided by numerical simulation of large graphs using tensor networks for $p<12$. Through simulation of worst-case graphs, which under the large loop conjecture have no small cycles, we compute the fixed angles that serve as the universally good parameters for any regular graph of connectivity 3, 4, or 5 and constant edge weight. The expectation value of these graphs serves as a performance guarantee for the QAOA, and are provided, along with angles, in Table \ref{table:3r_angles} and as a supplemental \texttt{JSON} file.

     For $p\geq 11$, the performance guarantee for QAOA on 3 regular graphs is larger than the guarantee of the Goemans Williamson algorithm, the best general-purpose MaxCut solver with a performance guarantee.
     This is a major step towards quantum advantage \cite{Bravyi2018advantage}. However, there exist algorithms that are more focused and provide better performance that GW, both guaranteed and heuristic. 
    
    As numerical evidence of the fixed angle conjecture, we evaluate the approximation ratio at fixed angles for all 3 regular graphs with $\leq 16$ vertices. We observe that no instance violates the performance guarantee, which proves that the fixed angle conjecture is \texttt{True} on all small graphs with $\leq 16$ vertices, and provides heuristic evidence that the conjecture is \texttt{True} for all 3 regular graphs.
    
    Additionally, we observe that the fixed angles are usually very close to global optima, with the typical difference between the fixed angle approximation ratio and global approximation ratio being $<0.003$ for $p=2$. This result is above and beyond the fixed angle conjecture, and removes the need for variational optimization by providing universal precomputed angles of Table \ref{table:3r_angles}. Similarly, we observe that with high likelihood that the fixed angles are usually ``close" to global optima, in that they are in the same basin of attraction. This may explain the previously observed phenomena of transfer of parameters  \cite{brandao2018fixed,streif2019training,galda2021transferability}, where good angles for one graph are good for others, because most graphs optimal angles are ``close" to these fixed angles.
    
    It is a curious fact that the optimal fixed angles look smooth, eg counterdiabatic \cite{wurtz2021counterdiabaticity}. In conjunction with the observation that global optima are very close to these fixed angles, it suggests that for a vast majority of graphs, smooth angles (under symmetry transformations in the parameter space of degenerate optima) may be optimal. This raises an interesting question: when are globally optimal angles adiabatic and smooth, vs non-adiabatic and non-smooth?
    
    In conclusion, we show that there is extra structure in the optimization landscape amongst ensembles of graphs, with fixed angles in parameter space being universally good among graphs. These fixed angles may allow QAOA implementations to bypass the optimization step and simply query one set of angles, speeding computation by potentially orders of magnitude. However, these results are limited in scope: the graphs must be $d$-regular, with fixed weights of $+1$ on every edge. An interesting future direction may be to evaluate fixed angle conjectures on more general graphs.
  
    While there are limits on graph degree and $p$ for which we are able to classically optimize the $\gb$ parameters, these results enable a single-shot QAOA on specific graphs. The power of QAOA may then reside only in the sampling capabilities of the quantum device.
    
    \section*{Acknowledgements}
	
	Numerical work was done on the Tufts High-Performance Computing Research Cluster
	and Argonne Leadership Computing Facility.
	JW and DL were supported by the Defense Advanced Research Projects Agency (DARPA) under Contract No. HR001120C0068, and JW was supported by NSF STAQ project (PHY-1818914).
	
	\bibliography{refs}
\end{document}